\documentclass[10pt]{article}
\usepackage{graphicx}
 \usepackage{amsfonts}
\usepackage{amssymb}
\usepackage{amsmath}
\usepackage{multicol}
\usepackage{color}
\usepackage{float}
\usepackage{verbatim}
\usepackage[utf8]{inputenc}
\usepackage[english]{babel}
\usepackage{amsfonts}
\usepackage{enumerate}
\usepackage{epsfig}

\begin{document}
\title{Modeling of X-ray attenuation via photon statistics evolution }
\author{Marc Guirao$^1$, Sergey Leble $^{1,2}$,\\[0.2cm]$^1$Faculty of Applied Physics and Applied Mathematics,\\Gdansk University of Technology,\\Narutowicza 11/12, 80-223, Gdansk, Poland \\$^2$Immanuele Kant Baltic Federal University,  \\Al. Nevsky st 41, Kaliningrad,  Russia}
\maketitle
\begin{abstract}
We consider a formulation of Cauchy problem for Kolmogorov equation
which corresponds a localized source of particles to be scattered by medium with given scattering amplitude density. The multiple scattering amplitudes are introduced and the corresponding series solution of the equation is constructed. We investigate the one-fold integral representation for the series terms, its estimations and values of photon number of a finite and point receivers. An application to X-ray beam scattering for orthogonal and  inclined to a layer  is considered.
\end{abstract}

\section{Introduction} This paper will be focused on a modeling of  X-ray propagation through a metal layer. Propagation of electromagnetic waves, such as X-ray, may be interpreted via two phenomena: absorption and scattering of the corresponding photons. Absorption of electromagnetic radiation is the way in which the energy of a photon is taken up by matter, typically the electrons of an atom. Thus, the electromagnetic energy is transformed into internal energy of the absorber, for example thermal energy. Scattering is the process in which moving particles or waves are forced to deviate from a straight trajectory by one or more paths due to localized non-uniformities in the medium through which they pass. Scattering phenomena can be divided in two classes: elastic and non-elastic scattering. In the elastic process the photon energy is conserved while in the non-elastic it is not. Rayleigh scattering is an example of elastic scattering. Non-elastic scattering has very low influence in X-ray propagation, therefore it will be ignored in this work. The simple model used in this theory to represent light-matter interaction is a crude assumption that avoids getting inside the actual complexity of the phenomena of light-matter interaction, and it has the advantage of leading to relatively simple analytical solutions for photon propagation valid for many real media. Light-matter interaction is characterized by several phenomena that will be neglected in the consideration since may to affect  the results of the investigations but could be taken into account by direct development of he theory. \\
\noindent
Scattering phenomena has been studied in Lidar (Light Detection and Ranging) problems since the 70's. Lidar works similar to radar. Laser shots short impulses of light in certain direction that is scattered by a medium (atmosphere \cite{Atmospheric} glass \cite{Attwood}, \cite{Hulst}, water \cite{Black_sea}, etc. \cite{Kerker}). Then the telescope collects the light that is scattered back and a gauge inside apparatus measures the intensity of that light. One of the first mathematical description are found in the works of \cite{Tomsk}. The backscattering phenomena is mathematically modeled by the linearized Boltzmann equation. that in fact is a version of Kolmogorov equations. Studies about the backscattering of a pulse emitted to the atmosphere through Monte Carlo simulations was done by \cite{Kunkel}, \cite{Moorodian}, and \cite{Spinhirne}. Double scattering was studied in the work \cite{BKaul}.  More recent work using Monte Carlo approach has been done by \cite{Miller} who studied the contribution of multiple scattering in the pulse stretching . Cloudy sky, fog or rain were considered for the simulations. A problem of mono energetic particles pulse reflection from half-infinite stratified medium is considered in conditions of elastic scattering with absorption account in the article \cite{Leble}. In this article the scattering phenomena is modeled by the Kolmogorov equation.
\noindent
Kolmogorov equations have their physical origin in Brownian Motion. The fact that Einstein  and Weiner  Brown motion models became obsolete increased attention to stochastic and probabilistic equations and resulted in Langevin, Smoluchowski and finally Fokker-Planck (Kolmogorov forward) equations. It is named after Adrian Fokker and Max Planck, German physicists who referred to it for the first time. It is also called from Andrey Kolmogorov who introduced it in a 1931 in his fundamental paper   "`Analytical
Methods in the Theory of Probability" some time after adding also the backward equation that was not known to Fokker and Planck. The first consistent microscopic derivation of the Fokker-Planck equation in the single scheme of classical and quantum mechanics was performed by Nikolay Bogoliubov and Nikolay Krylov. The second equation was called the backward equation from then. According to E. B. Dynkin, Kolmogorov was not familiar with the papers of Fokker and Planck \cite{Plank} in 1931 \cite{K1931}. Independently the geneticist R. A. Fisher studied the same equation \cite{Fok}.
 More recent articles on the LIDAR sounding of environment like \cite{Kaul} reveal interest to the related direct and inverse problems nowadays \cite{Uch} .\par
In this paper, a problem of mono-energetic X-ray particles pulse propagating in free space, reaching a layer of beryllium and going through free space again arriving to a cylindrical detector is considered. We simplify a bit the problem taking the speed of the electromagnetic waves will be constant through air and the beryllium layer. The theory is based on multiple scattering series solution of Kolmogorov equation for one-particle distribution function \cite{Leble} see also \cite{Risken}. Whereas in the articles cited the backscattering is considered, in this work we will focus on the forward scattering \cite{KolUch}. The main purpose of this paper is, by means of laboratory experimental data on the differential and total cross section \cite{Peng}, to obtain expressions for the intensity arriving to the detector after one-scattering phenomena \cite{Williams}. A point pulse source will be the initial condition.

\section{The problem formulation}

The equation for the probability density $f = f(t', \vec{r}, \vec{v})$ has the following form:
\begin{equation}\label{main}
\frac{1}{c}[\partial_{t'}f + \vec{v} \cdot \nabla f] = -\sigma_{tot}(z)f - \int \sigma_{scat}(\cos\gamma, z) f d\Omega' ,
\end{equation}
Where $t'$-time, $d\Omega'=\sin\theta'd\theta'd\phi'$ - solid scattering angle, $\sigma$- bulk differential cross-section of elastic scattering to the angle $\gamma$; $\sigma_{tot}$ - the sum of $\int \sigma_{scat} d\Omega + \sigma_{abs}$, scattering and absorption total cross-sections of elastic scattering, and 
\begin{equation}
\frac{\vec{v}}{c} = (\sin\theta \cos\phi, \sin \theta \sin \phi, \cos\theta)
\end{equation}
Equation \eqref{main} is derived from the Kolmogorov equation and has been taken from \cite{Leble}.
In spherical coordinates: $r, \theta, \phi$ the scattering angle is expressed as
\begin{equation}
\cos \gamma = \cos \theta \cos \theta' + \sin \theta \sin \theta' \cos(\phi-\phi').
\end{equation}
where $\theta'$ and $\phi'$ are the angles after the scattering occurs.
We suppose that the scattering is elastic, $|\vec{v}|$ does not change while the scattering process occur. Initial conditions are represented by distributions
\begin{equation}\label{icond1}
f(0,x,y,z) = V\delta(x)\delta(y)\delta(z)\delta(\theta-\theta_0).
\end{equation}
which means that an initial pulse is emitted from the position $(x,y,z) = (0,0,0)$ at a certain direction $\theta =\theta_0$. $V$ represents the number of photons emitted from the source. We will consider the parameter $V = \frac{1}{2\pi}$, so that the function $f$ can be interpreted as a density probability function.  

It means that we built a solution for the probability density as a weak limit (when $t' \rightarrow 0$) to $\delta-function$ at $t'>0$. The distribution $\delta (\theta-\theta_0)$ is chosen as 
\begin{equation}\label{delteta}
(\delta(\theta-\theta_0),\psi(\theta,\phi)) = \int_0^{2\pi} \psi(\theta_0,\phi) d\phi.
\end{equation}
The definition of the action of a  function $f$ on a function $\psi$ in $x,y,z$ coordinates from the Shwartz space is standard.
 
\section{Solution for the modeling equation}

\subsection{Solution for 0 angle initial pulse}
Let us start with a simple example of zero initial angle $\theta_0=0$. Denote $t = ct'$ and c the speed of light in air. This makes a unit of space and a unit of time equivalent. A solution is searched as an N-fold scattering expansion
\begin{equation}\label{series}
f = f_0 + f_1+f_2+...
\end{equation}
We are interested in 1-fold scattering, i.e, the approximation $f = f_0 + f_1$
We choose for $f_0$,
\begin{equation}\label{f0}
Lf_0 = \frac{\partial f_0}{\partial t}+ \sin\theta\cos\phi\frac{\partial f_0}{\partial x} + \sin\theta\sin\phi\frac{\partial f_0}{\partial y} + \cos \theta \frac{\partial f_0}{\partial z} = -\sigma_{tot}(z) f_0,
\end{equation}
and initial condition,
\begin{equation}
f_0(0,x,y,z) = V\delta(x)\delta(y)\delta(z)\delta(\theta).
\end{equation}
To find the equation for $f_1$ from \eqref{main} we write
$$Lf_0 + Lf_1 = -\sigma_{tot}(z)(f_0+f_1)-\int \sigma_{scat} (\cos \gamma, z) (f_0 + f_1) d\Omega'$$
We know from \eqref{f0} $Lf_0 = -\sigma_{tot}(z)f_0$, then
\begin{equation}\label{f1}
Lf_1 = -\sigma_{tot}(z)f_1 - \int \sigma_{scat} (\cos \gamma, z)f_0 d\Omega'
\end{equation}
with initial condition 
$$f_1|_{t=0} = 0.$$
Let us change variables in \eqref{f0} to solve the equation:
\begin{subequations}\label{change_variables}
  \begin{align}
    x' = x -t\sin\theta\cos\phi & \\
    y' = y - t\sin\theta\sin\phi & \\
    z' = z -t\cos \theta & \\
    t' = t &
  \end{align}
\end{subequations}
Therefore:
$$\dfrac{\partial}{\partial x} = \dfrac{\partial x'}{\partial x} \dfrac{\partial}{\partial x'} = \dfrac{\partial}{\partial x'}$$
$$\dfrac{\partial}{\partial y} = \dfrac{\partial y'}{\partial y} \dfrac{\partial}{\partial y'} = \dfrac{\partial}{\partial y'}$$
$$\dfrac{\partial}{\partial z} = \dfrac{\partial z'}{\partial z} \dfrac{\partial}{\partial z'} = \dfrac{\partial}{\partial z'}$$
$$\dfrac{\partial}{\partial t} = \dfrac{\partial x'}{\partial t} \dfrac{\partial}{\partial x'}+ \dfrac{\partial y'}{\partial t} \dfrac{\partial}{\partial y'} + \dfrac{\partial z'}{\partial t} \dfrac{\partial}{\partial z'} + \dfrac{\partial t'}{\partial t} \dfrac{\partial}{\partial t'} = -\sin\theta\cos\phi \dfrac{\partial}{\partial x'} - \sin\theta\sin\phi \dfrac{\partial}{\partial y'} -\cos \theta \dfrac{\partial}{\partial z'} + \dfrac{\partial}{\partial t'}$$
\noindent
The equation \eqref{f0} is transformed to 
\begin{equation}\label{transformed}
\frac{\partial f_0}{\partial t'} =  -\sigma_{tot}(z' + t' \cos \theta)f_0.
\end{equation}
$$f_0 = K(x', y', z') exp \left[ -\int_0^{t'} \sigma_{tot} (z' + \tau \cos \theta) d\tau \right]$$
Going back to the old variables:
$$f_0 = K(x-t\sin\theta\cos\phi, y - t\sin\theta\sin\phi, z -t\cos \theta) exp \left[-\int_0^t \sigma_{tot}(z - t \cos \theta + \tau \cos \theta) d\tau \right] $$
Let us remark a useful fact
\begin{equation}\label{simplification1}
exp \left[ -\int_0^t \sigma_{tot}(z - (t - \tau)\cos\theta))d\tau \right]. = exp \left[ -\int_0^t \sigma_{tot}(z - \tau\cos\theta))d\tau \right].
\end{equation}
Denote a function $E$ via 
\begin{equation}\label{E}
E(t,z,\theta) = exp \left[ -\int_0^t\sigma_{tot}(z-\tau\cos\theta)d\tau \right],
\end{equation}
Using the initial condition \eqref{icond1} with $V = \dfrac{1}{2\pi}$, the expression for $f_0$ is:
$$f_0 = \frac{1}{2\pi} \delta(x-t\sin\theta\cos\phi)\delta(y-t\sin\theta\sin\phi)\delta(z-t\cos\theta)\delta(\theta)$$
$$ \exp \left[ -\int_0^t \sigma_{tot}(z - (t-\tau) \cos\theta)d\tau \right].$$
Which, taking in account that the expression will not vanish only if $\theta = 0$, simplifies as:
\begin{equation}
f_0 = \frac{1}{2\pi} \delta(x)\delta(y)\delta(z-t)\delta(\theta) \exp \left[ -\int_0^t \sigma_{tot}(z - (t - \tau))d\tau \right].
\end{equation}
\begin{equation}\label{f0_solution}
f_0 = \frac{1}{2\pi} \delta(x)\delta(y)\delta(z-t)\delta(\theta)E(t,z,0). 
\end{equation}
This expression would be slightly different if the change of speed of light in different media would be taken in account.  Now, once we know $f_0$, we have to solve equation \eqref{f1}. We use the same change of variables \eqref{change_variables} to transform equation \eqref{f1} to 
\begin{align*}
f_1 e^{\int_0^{t'}\sigma_{tot}(z'+ \tau_2 \cos \theta)d\tau_2}  &= \\ 
 - \int_0^{t'} e^{\int_0^{\tau}\sigma_{tot}(z'+ \tau_2 \cos \theta)d\tau_2} \int\sigma_{scat}(\cos\gamma, z'+ \tau\cos\theta)f_0(\tau,x',y',z',\theta) d\Omega' d\tau\\ +  C_1
\end{align*}
\noindent
From initial conditions we conclude $C_1 = 0$,

\begin{align*}
f_1 &= V\int_0^{t'} e^{-\int_{\tau}^{t'}\sigma_{tot}(z'+ \tau_2 \cos \theta)d\tau_2} \int\sigma_{scat}(\cos \theta' , z'+ \tau\cos\theta)  \\
& E(\tau,  z' + \tau \cos \theta, 0)  \delta (x' + \tau \sin \theta \cos \phi) \delta (y' + \tau \sin \theta \sin \phi) \delta(z' + \tau \cos \theta - \tau) \delta(\theta')d\Omega'd\tau 
\end{align*}
Transformation to original variables, taking into account \eqref{delteta}, 
\begin{align*}
f_1 &= 2\pi\int_0^t e^{-\int_{\tau}^{t}\sigma_{tot}(z -(t - \tau_2) \cos \theta)d\tau_2} E(\tau,  z -(t - \tau) \cos \theta, 0)\sigma_{scat}(\cos \theta , z -(t-\tau)\cos\theta)  \\
&  V \delta (x -(t-\tau)\sin \theta \cos \phi) \delta (y -(t-\tau)  \sin \theta \sin \phi) \delta(z -(t-\tau) \cos \theta - \tau) d\tau, 
\end{align*}
 after simplification 

\begin{align*}
f_1 &= \int_0^t E(\tau, z, \theta) E(t-\tau,  z - \tau \cos \theta, 0)\sigma_{scat}(\cos \theta , z -\tau\cos\theta) \\
&  V \delta (x -\tau\sin \theta \cos \phi) \delta (y -\tau  \sin \theta \sin \phi) \delta(z - \tau \cos \theta - (t- \tau)) d\tau 
\end{align*}
\noindent
Integrations by $\theta, \phi, \tau$ are understood as integrations of the distribution by these parameters. For example, by the definition \eqref{action} $f_1$ acts on a function $\psi$ from Schwartz space as 
\begin{align}
(f_1(t,x,y,z,\theta,\phi),\psi(x,y,z))  &= \int_0^t E(\tau, \tau\cos\theta +t -  \tau, \theta) E(t-\tau, t-\tau, 0) \sigma(\cos\theta, t-\tau) \nonumber \\
&\qquad \psi(\tau\sin\theta\cos\phi, \tau\sin\theta\sin\phi, \tau\cos\theta + t - \tau) d\tau.
\end{align}
The function $\psi$ will be used to determine the position, size and configuration of the receiver.

\subsection{Initial condition for non zero angle initial pulse}

In the problem formulation we considered a pulse emitted from the source in the direction $\theta = 0$. 
We will also discuss the problem with an initial condition for different small angles of initial direction of $\theta = \theta_0$ of the pulse:
\begin{equation}\label{icond2}
f(0,x,y,z) = V\delta(x)\delta(y)\delta(z)\delta(\theta - \theta_0).
\end{equation}

\subsection{Solution for non zero angle initial pulse}

Proceeding the same way as before we get that the solution of $f_0$ with initial condition 
\begin{equation*}
f(0,x,y,z) = V\delta(x)\delta(y)\delta(z)\delta(\theta - \theta_0).
\end{equation*}
We take the general solution for a probability - normalized $f_0$
\begin{align*}
f_0 =& \frac{1}{2\pi} \delta(x-t\sin\theta\cos\phi)\delta(y-t\sin\theta\sin\phi)\delta(z-t\cos\theta)\delta(\theta - \theta_0)\\
& exp \left[ -\int_0^t \sigma_{tot}(z - (t-\tau) \cos \theta)d\tau \right],
\end{align*}
Which taking in account \eqref{E} and that the expression will not vanish only if $\theta = \theta_0$, simplifies:
\begin{equation}\label{f0no}
f_0 = \frac{1}{2\pi} \delta(x-t\sin\theta_0\cos\phi)\delta(y-t\sin\theta_0\sin\phi)\delta(z-t\cos\theta_0)\delta(\theta - \theta_0) E(t, z, \theta_0),
\end{equation}
or, going back to $\theta_0=0,$ yields
 \begin{equation}\label{f00}
f_0 = \frac{1}{2\pi} \delta(x)\delta(y)\delta(z-t)\delta(\theta)E(t,z,0). 
\end{equation}
The equation for $f_1$ in transformed variables \eqref{change_variables}
\begin{equation}\label{f11}
\begin{array}{c}
 f_{1t'} = -\sigma_{tot}(z'+t'cos\theta)f_1 - \\ \int \sigma_{scat} (\cos \gamma, z'+t'cos\theta)f_0(t',x'+t'sin \theta cos\phi,y'+t'sin\theta sin\phi, z'+t'cos\theta,\theta) d\Omega',
  \end{array}
\end{equation}
with initial condition 
$$f_1|_{t=0} = 0.$$
Integrating yields
\begin{equation}\label{f111}
\begin{array}{c}
 f_{1}e^{\int^{t'} _0\sigma_{tot}(z'+\tau'cos\theta)d\tau'} =  -\int_0^{t'}e^{\int^{\tau} _0\sigma_{tot}(z'+\tau'cos\theta)d\tau'} \\ \int \sigma_{scat} (\cos \gamma, z'+\tau cos\theta)f_0(\tau,x'+\tau sin \theta cos\phi,y'+\tau sin\theta sin\phi, z'+\tau cos\theta,\theta) d\Omega'd\tau  
 \end{array}
\end{equation}
The solution   $f_1$ for this initial conditions in original variables is 
\begin{equation}\label{f112}
\begin{array}{c}
 f_{1}  = - E(t,z,\theta) \int_0^te^{\int^t _{t-\tau}\sigma_{tot}(z-\tau''cos\theta)d\tau''}   \int \sigma_{scat} (\cos \gamma, z+(t-\tau) cos\theta) \\ f_0(\tau,x+(t-\tau) sin \theta cos\phi,y'+(t-\tau) sin\theta sin\phi, z-(t-\tau) cos\theta,\theta) d\Omega'd\tau.
 \end{array}
\end{equation}
  Changing variables of integration $t-\tau=\tau'$ and omitting the primes,
\begin{equation}\label{f1}
\begin{array}{c}
	f_1 = -  \int_0^t E(\tau,z,\theta)  
	\int\sigma_{scat}(\cos \theta \cos \theta' + \sin \theta \sin \theta' \cos(\phi-\phi'), z-\tau\cos\theta)\\f_0 (t-\tau, x-\tau\sin\theta\cos\phi,y-\tau\sin\theta\sin\phi,z-\tau\cos\theta,\theta)d\Omega'd\tau
\end{array}
\end{equation}
Plugging point pulse \eqref{f0no} yields
\begin{equation}
\begin{array}{c}
	f_1 = -  \frac{1}{2\pi}\int_0^t E(\tau,z,\theta) E(t-\tau,z-\tau\cos\theta,\theta)
	\\ \int\sigma_{scat}(\cos \theta \cos \theta' + \sin \theta \sin \theta' \cos(\phi-\phi'), z-\tau\cos\theta)\\\frac{1}{2\pi} \delta(x-\tau\sin\theta\cos\phi)\delta(y-\tau\sin\theta\sin\phi)\delta(z-\tau\cos\theta)\delta(\theta - \theta_0)d\Omega'd\tau
\end{array}%
\end{equation}
or, taking into account independence of the indicatrix $\sigma_{scat}$  on $\phi'$ and $\delta(\theta - \theta_0)$ definition,
\begin{equation}
\begin{array}{c}
	f_1 =  -\int_0^t E(\tau,z,\theta_0) E(t-\tau,z-\tau\cos\theta_0,\theta_0)
	\int_0^{\pi}\sigma_{scat}(\cos \theta_0 \cos \theta'  , z-\tau\cos\theta_0)\\  \delta(x-\tau\sin\theta_0\cos\phi)\delta(y-\tau\sin\theta_0\sin\phi)\delta(z-\tau\cos\theta_0)\delta(\theta - \theta_0)\sin\theta'd\theta'd\tau
\end{array}%
\end{equation}
The distribution action for the cylindrical symmetry  is evaluated as
\begin{equation}\label{act}
\begin{array}{c}
	(f_1,\psi) =- \int_0^t E(\tau,z,\theta_0) E(t-\tau,z-\tau\cos\theta_0,\theta_0)
	\int_0^{\pi}\sigma_{scat}(\cos \theta_0 \cos \theta'  , z-\tau\cos\theta_0)\\  \psi(\tau\sin\theta_0\cos\phi ,\tau\sin\theta_0\sin\phi 
	,\tau\cos\theta_0)\sin\theta'd\theta'd\tau.
\end{array}%
\end{equation}

\section{Number of particles rate}
\subsection{Number of particles rate for $\theta_0$ angle initial pulse}

Generally \cite{Tomsk}, the probabilistic  interpretation of the distribution function $f$ in the phase space gives the number of particles in a small volume $\Delta x \Delta y\Delta z$ as
$$
\int_0^{2\pi}\int_{\pi-\theta_0}^{\pi}\int_x^{x+\Delta x}\int_y^{y+\Delta y}\int_z^{z+\Delta z} f dxdydz \sin\theta d\theta d\phi ,
$$
For a point receiver at $x,y,z$ it is found as a limit 
\begin{equation}\label{I}
	I(t,x,y,z)=\lim_{\Delta x \rightarrow 0, \Delta y\rightarrow 0,\Delta z\rightarrow 0}\int_0^{2\pi}\int_{\pi-\theta_0}^{\pi}\int_x^{x+\Delta x}\int_y^{y+\Delta y}\int_z^{z+\Delta z} f dxdydz \sin\theta d\theta d\phi ,
\end{equation}
here an aperture angle $\theta_0$, that restricts possible velocities of particles directions is introduced. 

Whereas in the Lidar problem the receiver is placed at the origin \cite{Tomsk}, in our problem it is placed at a certain distance after the scatterer layer. We will place the receiver at the position $(x,y,z) = (0,0,z_0)$ and the layer at $z \in [\dfrac{z_0}{2} - \Delta, \dfrac{z_0}{2} +\Delta] $, being $2\Delta$ the thickness of the layer. Our aim is the evaluation of number of particles which enter the round area of radius $\rho_0$ laying in the plane $z = z_0$ (receiver) with center in the origin and having velocity vectors inclined to z-axis within the angle interval $\theta \in [0, \alpha]$. The angle $\alpha$ relates the aperture angle of a receiver. By its direct sense, the number is proportional to
the number of particles (photons) per unit time and volume is given by general relation
\begin{equation}\label{Intensity}
I(t) = \lim\limits_{\Delta t \rightarrow 0} \frac{1}{\Delta t} \int_{0}^{\alpha}\int_0^{2\pi}(f(t,0,0,z_0,\theta,\phi),\psi(x,y,z,\theta,\phi))\sin\theta d\phi d\theta.
\end{equation} 
The choice of the function $\psi$ can be realized by concrete physical reasons. In the exemplary case we take here, the receiver has cylindrical symmetry and for the initial direction along $z$, the function does not depend on $\theta, \phi $, so it is chosen zero outside the receiver, and  $\psi(x,y,z) = 1$ for internal points of the domain $x^2 + y^2 \leq \rho_0^2$, \qquad $z_0 \leq z \leq z_0 + \Delta t |\cos \theta|$ and zero outside, being $z_0$ the coordinate and $|z_0|$ of the distance between the source of the pulse and the receiver.
Evaluation of 1-fold scattering by \eqref{act} for a point receiver we do as the limit:
\begin{align*}
I_1(t) = - \lim\limits_{\Delta t \rightarrow 0} \lim\limits_{\rho_0 \rightarrow 0} \frac{1}{\Delta t} \int_0^{\rho_0}\int_0^{\alpha} \int_0^t E(\tau,z_0,\theta_0)  E(t-\tau,z-\tau\cos\theta_0,\theta_0)\\
	\sigma_{scat}(\cos \theta_0 \cos \theta'  , z-\tau\cos\theta_0)  \psi(\tau\sin\theta_0\cos\phi ,\tau\sin\theta_0\sin\phi 
	,\tau\cos\theta_0)\sin\theta'd\theta'd\tau d\phi d\rho.
\end{align*}
 Therefore, we get for \eqref{f1}
\begin{equation}\label{f1I11}
\begin{array}{c}
	I_1 (t,0,0,z_0)= -  \lim\limits_{\Delta t \rightarrow 0} \lim\limits_{\rho_0 \rightarrow 0}\int_0^{2\pi}\int_0^{\rho_0}\int_0^{\alpha}\int_0^t  E(\tau,z_0,\theta)
	\int\sigma_{scat}(\cos \theta \cos \theta'  , z-\tau\cos\theta)\\f_0 (t-\tau, x-\tau\sin\theta\cos\phi,y-\tau\sin\theta\sin\phi,z-\tau\cos\theta,\theta)\sin\theta'd\theta'd\tau d\rho d\phi
\end{array}
\end{equation}
 The nonzero $I_1$ values are obtained in conditions that are explained by means of the figure \ref{fig:plot1} below.
 
\begin{figure}[h]
	\centering
		\includegraphics{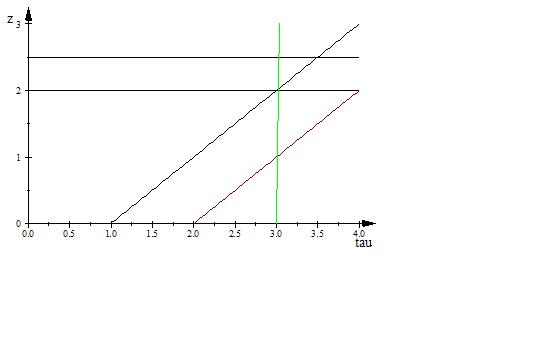}
	\label{fig:plot1}
		\caption{The figure explains conditions of nonzero contribution into intensity.}
\end{figure}

	 The area of integration lies between the horizontal lines $z=z_0,z_0+\Delta z$ and inclined lines $z=\cos\theta_0+a$ and $z=\cos\theta_0+b$, where a,b are boundaries of a "cloud" or metal plate. The vertical green line marks a pulse arrival time $z=t$.

It would be convenient to change variables  Taking in account the definition of the integration of a. distribution function $\delta$, we introduce  new variables  by the following equalities:
\begin{align}\label{xyz}
x' &= \tau \sin \theta \cos \phi \nonumber\\
y' &= \tau \sin\theta \sin \phi \nonumber \\
z' &= \tau \cos \theta + t - \tau
\end{align}
\noindent
the inverse one are found as
\begin{align}\label{inverse}
\tau &= \dfrac{x'^2+y'^2+z'^2+t^2-2tz'}{2(t-z')} \nonumber \\
\cos \theta &=  \dfrac{x'^2+y'^2+z'^2+t^2-2tz' -2(t-z')^2}{x'^2+y'^2+z'^2+t^2-2tz'} \nonumber \\
\cos\phi &= \dfrac{x'}{\sqrt{x'^2+y'^2}},\\
\tau\cos \theta &= \dfrac{\rho^2 -(t-z')^2}{2(t-z')}= \dfrac{(\rho -(t-z'))(\rho+(t-z'))}{2(t-z')}.
\end{align}
\noindent
The Jacobian $J$ is evaluated as 
\begin{equation*} 
J^{-1} = det 
\begin{vmatrix}
\frac{dx}{d\tau} & \frac{dx}{d\theta} & \frac{dx}{d\phi} \\
\frac{dy}{d\tau} & \frac{dy}{d\theta} & \frac{dy}{d\phi} \\
\frac{dz}{d\tau} & \frac{dz}{d\theta} & \frac{dz}{d\phi} 
\end{vmatrix}
= \dfrac{x'^2+y'^2+z'^2+t^2-2tz'}{2} \sin\theta
\end{equation*}

\begin{equation*}
J \cdot \sin \theta = \dfrac{2}{x'^2+y'^2+z'^2+t^2-2tz'}
\end{equation*}
Let's define now the integration interval limits
\begin{align}\label{interval_limits}
0 \leq \tau \leq t \nonumber \\
0 \leq \phi \leq 2\pi \nonumber \\
0 \leq \theta \leq \alpha \nonumber \\
z_0 \leq \tau\cos\theta + t - \tau \leq z_0 + \Delta z \nonumber \\
0 \leq \rho^2=\tau^2\sin^2\theta \leq \rho_0^2 \nonumber \\
\end{align}

 Hence, the formula \eqref{f1I11} for the new variables is
\begin{equation}
\begin{array}{c}
	I_1 (t,x,y,z)= -  \lim\limits_{\Delta t \rightarrow 0}  \int_x^{x+\Delta x}\int_y^{y+\Delta y}\int_z^{z+\Delta z} E(\tau,z,\theta) 
	\int\sigma_{scat}(\cos \theta \cos \theta'  , z-\tau\cos\theta)\\ \sin\theta'd\theta' \dfrac{2f_0 (t-\tau, x-x',y-y',z-z'+t-\tau,\theta)}{x'^2+y'^2+z'^2+t^2-2tz'} dx'dy'dz'.
\end{array}
\end{equation}
Plugging $f_0$ gives 
\begin{equation}
\begin{array}{c}
I_1(t,x,y,z)=	  - \frac{1}{\pi} \lim\limits_{\Delta x, \Delta y, \Delta z \rightarrow 0} \frac{2}{\Delta x \Delta y \Delta z} \int_{0}^{\Delta x} \int_x^{x+\Delta x}\int_y^{y+\Delta y}\int_z^{z+\Delta z} E(\tau,z,\theta) 
\\	\int\sigma_{scat}(\cos \theta \cos \theta'  , z-\tau\cos\theta)\\ \sin\theta'd\theta' \dfrac{ \delta(x-x')\delta(y-y')\delta(z-z'+t-\tau)\delta(\theta - \theta_0) E(t-\tau, z, \theta_0) 
}{x'^2+y'^2+z'^2+t^2-2tz'} dx'dy'dz'.
\end{array}
\end{equation}
By a definition of delta-functions at $x\in [x,x+\Delta], ...$ taking into account $z'-t+\tau=\frac{x'^2+y'^2-(z'-t)^2}{2(t-z')}$ with two roots 
 \begin{equation}
\begin{array}{c}
I_1(t,x,y,z)=	  - \frac{1}{\pi} \lim\limits_{\Delta x, \Delta y, \Delta z \rightarrow 0} \frac{2}{\Delta x \Delta y \Delta z}\int_x^{x+\Delta x}\int_y^{y+\Delta y}\int_z^{z+\Delta z}
	\int\sigma_{scat}(\cos \theta \cos \theta', z-\tau\cos\theta)\\ \sin\theta'd\theta' \dfrac{   \delta(\theta - \theta_0) E(\tau,z,\theta) E(t-\tau, z, \theta_0) 
}{x^2+y^2+(z+t-\tau)^2+t^2-2t(z+t-\tau)}.
\end{array}
\end{equation}
Because of dealing with a point receiver and its position we should introduce limits in $\Delta x$ $\Delta y$ $\Delta z$.
or, for  the point receiver at $(0,0,z_0)$ and \eqref{f0no}
\begin{equation}
I_1(t) =I_1 (t,0,0,z_0)=\frac{1}{2} \lim\limits_{\Delta z \rightarrow 0} \frac{2}{\Delta z} \int_{z_0}^{z_0 + \Delta z}  \dfrac{E' E''\sigma_{scat}'}{ z^{'2} + t^2 -2tz'}\delta(2z+t-z' ) dz' 
\end{equation}
Where $E'$ , $E''$ , $\sigma_{scat}'=\sigma_{scat}(\cos\gamma,t-\dfrac{z'^2+t^2-2tz'}{2(t-z')} )$ are elements of \eqref{f0no} in new variables \eqref{inverse}, plugging $f_0$ 
 results in the compact formula
 
\begin{equation}\label{I1}
\begin{array}{c}
I_1(t)=	I_1 (t,0,0,z_0)= -     
	\int\sigma_{scat}(\cos \theta_0 \cos \theta', z_0+t-\tau)\\ \sin\theta'd\theta' \dfrac{ E( \frac{t-z_0}{2},z_0,\theta_0)  E(z_0, z_0, \theta_0) 
}{ (2z+t)^2+t^2-2t(2z+t)}  .
\end{array}
\end{equation}
The result for zero angle for the point receiver is almost trivial from geometrical point of view, the arriving pulse is infinitely short. The expression for intensity \eqref{I1}  contains natural spherical divergence, exponential decay due to absorption and forward scattering i n a level inside the layer.

\subsection{Modeling $\sigma_{scat}$ and $\sigma_{tot}$ from experimental data}
Let's consider now the scattering in homogeneous layer of a given material. You can better understand the situation by looking at the picture.
\begin{figure}[H]
\begin{center}
\includegraphics[keepaspectratio=true,scale=0.1]{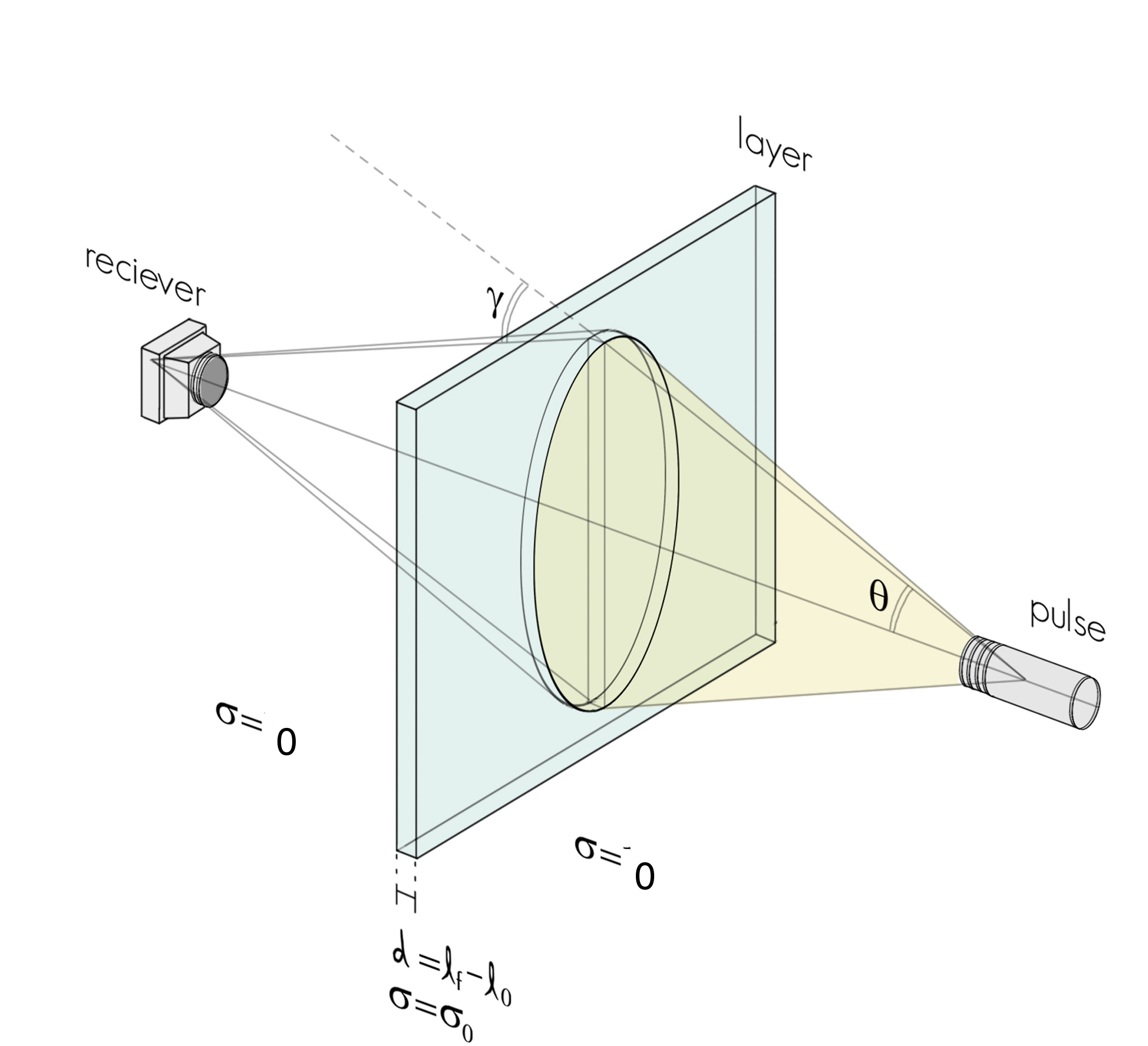}
\caption{Situation modeled. $\sigma = \sigma(1,z)$.}
 \end{center}
 \end{figure}
Depending on the X-ray energy and the material of the layer, we will have different functions for $\sigma_{scat}$. First, let's consider a simple case. 
For $ \cos \gamma = 1$
\begin{displaymath}\label{sigma1}
   \sigma_{scat}(\cos \gamma, z) = \left\{
     \begin{array}{lr}
       0,   & z \leq \frac{z_0}{2} - \Delta \\
       \sigma_0,   & \frac{z_0}{2} - \Delta < z < \frac{z_0}{2} + \Delta \\
       0,   & z \geq \frac{z_0}{2} + \Delta
     \end{array}
   \right.
\end{displaymath} 
%
%
%
\begin{figure}[H]
\begin{center}
\includegraphics[keepaspectratio=true,scale=0.3]{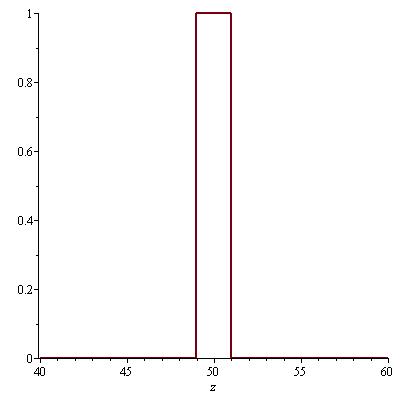}
\end{center}
\caption{Proposed $\sigma_{scat}(1,z)$.}
\end{figure}
\noindent
\begin{displaymath}\label{sigmat}
   \sigma_{tot}(z) = \left\{
     \begin{array}{lr}
       0,   & z\leq \frac{z_0}{2} - \Delta\\
       \sigma_{1},   & \frac{z_0}{2} - \Delta < z < \frac{z_0}{2} + \Delta \\
       0,   & z \geq \frac{z_0}{2} + \Delta
     \end{array}
   \right.
\end{displaymath} 
\noindent
The total attenuation cross section $\sigma_{tot}$ can be divided in two terms: scattering total cross section $\int_0^{\pi} \sigma_{scat}$ and absorption total cross section $\sigma_{abs}$.
To model $\sigma_{tot} = \int_0^{\pi} \sigma_{scat}(\cos \gamma, z) d\gamma + \sigma_{abs}$, let's first consider some theory. 
The \textbf{Attenuation Lambert-Beer Law} states;
$$I = I_0 e^{-\mu z}$$
where $\mu$ is the attenuation coefficient. 
In the next figure we can see how the intensity of the X-ray beam decreases as it penetrates a layer of beryllium.
\begin{figure}[H]
\begin{center}
\includegraphics[keepaspectratio=true,scale=0.3]{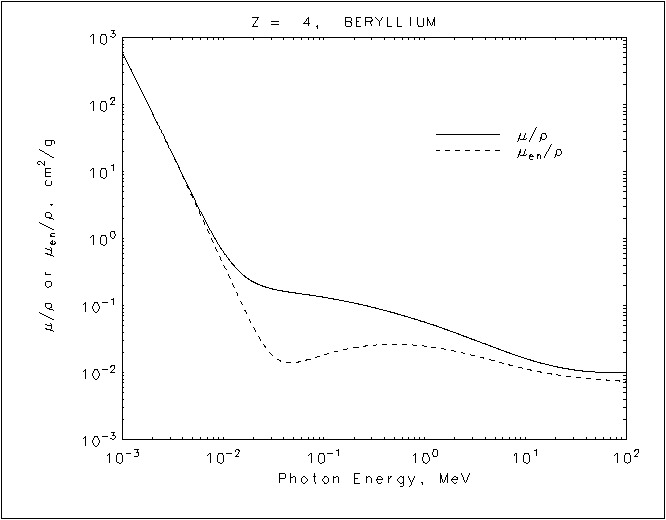}
\end{center}
\caption{Beryllium intensity attenuation.}
\end{figure}
\noindent
Now, $\mu$ is defined as
\begin{equation}\label{attenuation}
\mu = \rho_a \sigma_{tot} = \rho \frac{N_0}{A}\sigma_{tot}
\end{equation}
where $\rho_{atom} = \rho \frac{N_0}{A}$ is the atomic density, $N_0$ is the Avogadro's number, $A$ is the atomic mass number and $\rho$ is the density ($g/m^3$).
\begin{equation}\label{absorption}
\sigma_{tot} =  \frac{A\mu}{\rho N_0}
\end{equation}
On the other hand, the Rayleigh scattering cross section $\sigma_{scat}$ expression taken from \cite{International_table} is
\begin{equation}\label{Rayleigh}
\sigma_{scat} = \pi r_e^2 \int_{-1}^1 (1+\cos^2\gamma) f^2(q,Z) d(\cos \gamma)
\end{equation}
where 
$r_e$ is the electron radius,\\
$\gamma$ is the scattering angle,\\
$2\pi d(\cos \gamma)$ is the solid angle between cones with angles $\gamma$ and $\gamma + d \gamma$,\\
$f(q,Z)$ is the atomic scattering factor,\\
$q$ is $\frac{\sin \gamma /2}{\lambda}$, the momentum transfer parameter and \\
$\lambda$ is the wavelength expressed in \AA.\\
Next you can see the plots of the experimental data for the atomic scattering factor $f(q,Z)$ for beryllium.
\noindent
\begin{figure}[H]
\begin{center}
\includegraphics[keepaspectratio=true,scale=0.3]{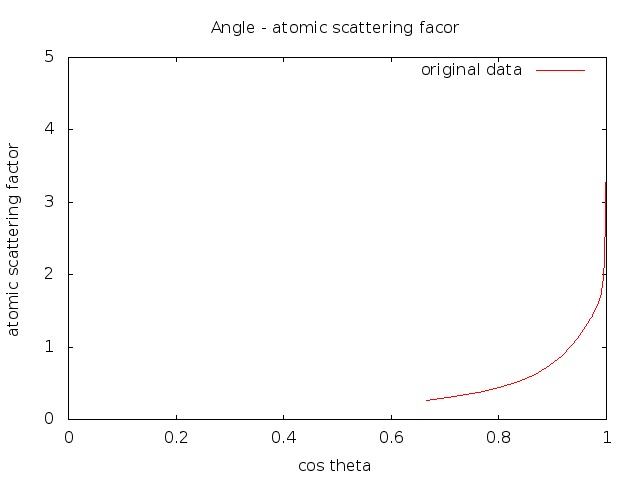}
\end{center}
\caption{Atomic scattering factor for beryllium.}
\label{atomic_scattering}
\end{figure}
We will consider the values for the total photon interaction cross section for beryllium from the tables of \cite{International_table}. We are interested in beryllium due to its properties and feasible applications. The values of table consider also the Compton scattering, i.e, $\sigma_{tot} = \sigma_{Rayleigh} + \sigma_{Compton} + \sigma_a $. However, we will not talk about the Compton scattering as it is sometimes neglected due to its low influence in the X-ray scattering.
\begin{table}[H]
\centering
    \begin{tabular}{ | l | l | l |}
    \hline
    Radiation & Energy(KeV) & $\sigma_{atom}$ $(\frac{barns}{atom})$ \\ \hline
    Ag $K\beta_1$ & 24,94 & 2,97  \\ \hline
    Zn $K_\alpha^-$ & 8,63 & 13,8   \\ \hline
    Mn  $K_\alpha^-$ & 5,895 & 39,9 \\
    \hline
    \end{tabular}
    \caption{Scattering for Be.}\label{table1}
\end{table}
\noindent
As we see in the table the total cross section is given in units $barns/atom$. We should normalize it per unit volume, i.e, $cm^{-1}$.\\
\\
\begin{center}
$\sigma_{tot} = N \sigma_{atom}$ \qquad where $ N = \dfrac{\rho N_0}{A}$.\\
$\left[cm^{-1} \right] = \sigma_{tot} = \dfrac{\rho N_0 \sigma_{atom} 10^{-22}}{A} = \left[ \dfrac{\dfrac{atom}{mol}\dfrac{g}{cm^3}\dfrac{cm^2}{atom}}{\dfrac{g}{mol}}\right] =\left[ cm^{-1} \right].$
\end{center}
For beryllium $\rho = 1,85 \frac{g}{cm^3}$, $A = 9,01$ and $\sigma_{atom} = 2,97$ for a 24,94 KeV energy radiation.

\subsection{Intensity plots}

The following plots will be made taking in account a layer of beryllium of thickness $2\Delta$ in the half way from the source to the receiver, i.e, at the range $\frac{z_0}{2} - \Delta \leq z \leq \frac{z_0}{2} + \Delta$. We will plot the results of the 1-scattering approach with the $0$ angle initial pulse. The intensity plot comes from formula \eqref{formula1}. After simplification of \eqref{formula1}:
\begin{equation*}
I(t) = \frac{2}{(t - z_0)^2} exp\left[-\int_0^t \sigma_{tot}(\tau)d\tau - 2\int_t^{\frac{t+z_0}{2}} \sigma_{tot}(\tau)d\tau \right]\sigma_{scat}(1, \frac{t+z_0}{2}).
\end{equation*}
\noindent
with the variable values defined as $z_0 = 100 m$, the distance from the source to the receiver, 

\begin{displaymath}
   \sigma_{tot}(z) = \left\{
     \begin{array}{lr}
       0,   & z \leq  \frac{z_0}{2} - \Delta\\
       \dfrac{\rho N_0}{A} \sigma_{atom},   & \frac{z_0}{2} - \Delta < z < \frac{z_0}{2} + \Delta \\
       0,   & z \geq \frac{z_0}{2} + \Delta
     \end{array}
   \right.
\end{displaymath} 
, $\rho = 1,85 \frac{g}{cm^3}$, the density of beryllium, $A = 9,01$ the beryllium atomic mass number, $\sigma_{atom} = 2.97$ and $N_0$ Avogadro's number. For the scattering cross section we used the fitted function to the atomic scattering factor of beryllium from figure \ref{atomic_scattering}. We can see both data and fitted function in the next figure
\begin{displaymath}
   \sigma_{scat}(\cos \gamma, {z}) = \left\{
     \begin{array}{lr}
       e^{1,2238 (\cos\gamma)^8} - 1 & : z \in (\frac{z_0}{2} - \Delta, \frac{z_0}{2} + \Delta) \\
       0 & : z \notin (\frac{z_0}{2} - \Delta, \frac{z_0}{2} + \Delta)
     \end{array}
   \right.
\end{displaymath} 
\begin{figure}[H]
\begin{center}
\includegraphics[keepaspectratio=true,scale=0.3]{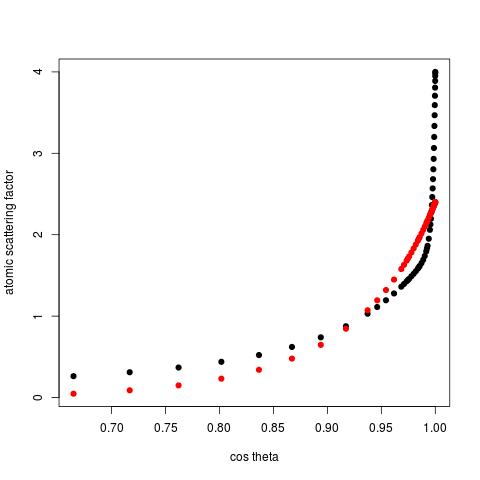}
\end{center}
\caption{In black the data function. In red the fitted function.}
\end{figure}
 
\begin{figure}[H]\label{fig:ris}
   \includegraphics[keepaspectratio=true,scale=0.5]{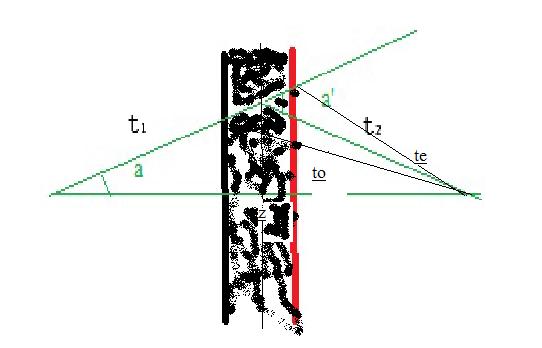}
\caption{The trajectories of a scattered photons from minimal to maximum scattering angles.}
\end{figure}

 To plot the intensities we will consider the thickness of the beryllium layer $2\Delta$. 
 The resulting formula for the point receiver intensity has direct geometrical interpretation (see Fig\ref{fig:ris}) 
 As it is marked in the plot the total time for a photon arrival is $t=t_{1}+t_{2}=\frac{z}{\cos \vartheta _{0}}+\frac{z_{0}-z}{\cos \beta }$

The angle is evaluated via

$\tan \beta =\frac{z\tan \vartheta _{0}}{z_{0}-z\ },$

and

$\cos \beta =\sqrt{\frac{1}{1+\tan ^{2}\beta }}=\sqrt{\frac{1}{1+(\frac{%
z\tan \vartheta _{0}}{z_{0}-z\ })^{2}}}=\sqrt{\frac{\left( z-z_{0}\right)
^{2}}{z^{2}\tan ^{2}\vartheta _{0}+z^{2}-2zz_{0}+z_{0}^{2}}}$

$t=\ \frac{z}{\cos \vartheta _{0}}+\frac{z_{0}-z}{\sqrt{\frac{\left(
z-z_{0}\right) ^{2}}{z^{2}\tan ^{2}\vartheta _{0}+z^{2}-2zz_{0}+z_{0}^{2}}}}$

 the expression of the scattering point height via time $t$ is
 
  $z=\frac{1}{2t\cos \vartheta _{0}-2z_{0}\cos ^{2}\vartheta _{0}}%
\left( t^{2}\cos ^{2}\vartheta _{0}-z_{0}^{2}\cos ^{2}\vartheta _{0}\right)
=\allowbreak \frac{1}{2}\left( \cos \vartheta _{0}\right) \left(
t-z_{0}\right) \frac{t+z_{0}}{t-z_{0}\cos \vartheta _{0}}$

with the correspondent $\cos \beta$ and $\sin \beta $ expressions (see the Appendix with SWP program.)
 Recall that
 
 $\cos \gamma =\cos \beta \cos \vartheta _{0}-\sin \beta \sin
\vartheta _{0}$

The relative intensity formula follows from \eqref{}

$\frac{I}{I_{0}}=\frac{1}{\left( t_{2}\right) ^{2}}\exp [-\sigma_{t}(t_{3}+t_{4})]\sigma (\cos \gamma ,z)$

where

$t_{3}=\frac{z-\frac{z_{0}}{2}+\Delta }{\cos \vartheta _{0}}$

$t_{4}=\frac{-z+\frac{z_{0}}{2}+\Delta }{\cos \beta }$

 the arrival time

$t_{0}=\ \left[ \frac{z}{\cos \vartheta _{0}}+\frac{z_{0}-z}{\sqrt{\frac{%
\left( z-z_{0}\right) ^{2}}{z^{2}\tan ^{2}\vartheta
_{0}+z^{2}-2zz_{0}+z_{0}^{2}}}}\right] _{z=\frac{z_{0}}{2}-\Delta
} $

and the end of the pulse is evaluated as 

\bigskip $t_{e}=\ \left[ \frac{z}{\cos \vartheta _{0}}+\frac{z_{0}-z}{\sqrt{%
\frac{\left( z-z_{0}\right) ^{2}}{z^{2}\tan ^{2}\vartheta
_{0}+z^{2}-2zz_{0}+z_{0}^{2}}}}\right] _{z=\frac{z_{0}}{2}+\Delta
} ,$

A convenient for illustration plotting  choice of parameters $\vartheta
_{0}=0.1,z_{0}=0.01,\Delta =0.001$ give $ t_{0}=3\times
10^{-2}, \,t_{0}=0075\times
10^{-2}  $.

The plot of has been made with SWP and is of the corresponding form.

\begin{figure}[h]
\begin{center}
\includegraphics[keepaspectratio=true,scale=0.7]{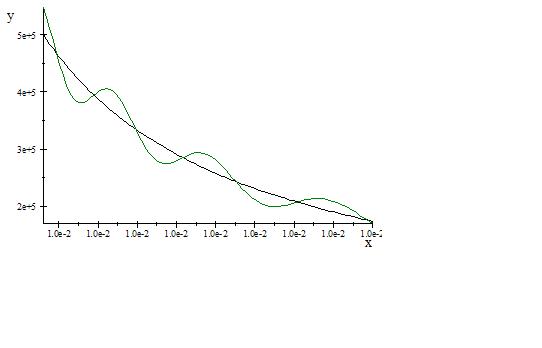}
\end{center}
\caption{Intensity for 1-scattering approximation, inclined beam. The green line is result of a superposition of the constant density with periodic one.
}
\end{figure}
\noindent
The beam has traveled from the source through air, where we do not account for scattering or absorption. It has penetrated the beryllium layer where the scattering and absorption phenomena have occurred. It has traveled through air again and finally has arrived to the receiver. We can see that the receiver will detect a delayed and spread pulse with a peak at time 100 and it will rapidly  decrease as the scattered photons arrive. It is a finite pulse as it is the traveling initial pulse.
\subsection*{Acknowledgment}
 The work is supported by Ministry of Education and Science of the Russian Federation (Contracts No 14.Y26.31.0002 and 02.G25.31.0086)"

\section{Conclusions}
We have obtained a solution of the Kolomogorov \cite{K1931} forward equation \cite{Fokker, Kolchuzhkin} for 1-fold scattering for an initial pulse of angle $\theta = \theta_0$. We have modeled the equations for the total scattering cross section  $\sigma_{tot}$ and the differential cross section $\sigma_{scat}$ for beryllium. Once we have this there has been derived an expression for the intensity rate from the initial pulse arriving at a small receiver situated after the beryllium layer at position $z=z_0$. Particularly, the results are obtained with a point receiver. After plotting the result of this intensity it has been shown that the pulse arrive to the receiver with certain delay and spread. With this formulas, considering other materials and modeling the scattering cross section of them, we are able to predict the delay and intensity of the initial pulse. 
There is a space for continuation of this work \cite{Pasc}. N-fold approximation is in our interest. The
next step is to calculate it using the recurrent relation \eqref{series}. When the f is calculated in
limit, we can count the stream for it. Properties of the layers can be obtained by studying the plots from the receiver. There are many situations that could be studied in the future. It is a matter of changing the initial condition. To be more realistic, the initial pulse should be taken continuous in time. It can be studied an initial continuous pulse distributed in a solid angle, $\theta \in [\theta_0, \theta_1]$ $\phi \in [\phi_0, \phi_1]$. Different layer materials could be used as well as the position and thickness of the layer. A very interesting approach would be considering heterogeneities inside the layer materials. In this case it should be introduced a distribution of the heterogeneities in the media. This would be more realistic, as materials present non-homogeneities in their structures.

\end{document}